\documentclass[final,5p,times,twocolumn]{elsarticle}  

\usepackage{lineno,hyperref}
\modulolinenumbers[5]

\journal{arXiv}

\usepackage{amsmath}
\usepackage{amssymb}

\begin{document}

\begin{frontmatter}

\title{Characterization of sealed (zero gas flow) RPCs under strong irradiation source}

\author[lipAddress]{A. Blanco}
\author[depAddress,lipAddress]{Mário Carvalho}
\author[lipAddress,isecAddress]{Susete Fetal}
\author[lipAddress]{Luis Lopes}
\author[depAddress,telmoAddress,lipAddress]{Telmo Paes}
\author[depAddress,lipAddress]{Joana Pinto}
\author[lipAddress,isecAddress]{Paulo Fonte}

\address[lipAddress]{LIP, Laborat\'{o}rio de Instrumenta\c{c}\~{a}o e F\'{i}sica Experimental de Part\'{i}culas, 3004-516 Coimbra, Portugal}
\address[depAddress]{Departamento de F\'{i}sica Universidade de Coimbra, 3004-516, Coimbra, Portugal}
\address[isecAddress]{Instituto Polit\'ecnico de Coimbra, Instituto Superior de Engenharia de Coimbra, Rua Pedro Nunes, 3030-199 Coimbra, Portugal}
\address[telmoAddress]{Instituto Federal do Esp\'{i}rito Santo, Cariacica, Brazil}

\begin{abstract}
The phase-out of Hydrofluorocarbons (HFCs), due to their high Global Warming Potential (GWP), affecting the main gas used in Resistive Plate Chambers (RPCs), tetrafluorethane C$_2$H$_2$F$_4$, has created operational difficulties on existing systems and imposes strict restrictions on its use in new systems. A possible solution to the problem is the substitution of this gas by others with a much lower GWP. This approach has attracted the attention of an important part of the community in recent years.

But there could be another possibility, which is the construction of sealed chambers, i.e. chambers that do not need a continuous gas flow to operate (in a similar way to Geiger-M{\"u}ller counters). This possibility would allow continuing to use HFCs or eventually other types of gases that are already banned. It would also greatly simplify the detector, not requiring the complex and expensive gas systems normally used. This simplification could be particularly relevant for outdoor applications, such as in Cosmic Ray experiments. This has motivated a new line of research and development on sealed RPCs: RPCs that do not require a continuous gas flow for their operation.

In this work we show the characterization of a sealed RPC plane exposed to an intense radiation source for about one month, with the aim of verifying, in an accelerated way, gas degradation effects that could compromise the operation of the detector. The results show no apparent degradation in any of the parameters studied, namely: efficiency, average charge and streamer probability. The exposure period corresponds, based on the accumulated charge, to approximately four years of natural Cosmic Ray radiation exposure (background signals generated by the detector itself), suggesting the feasibility of the application of this technology in Cosmic Ray experiments for extended periods of time.
\end{abstract}

\begin{keyword}
Resistive Plate Chambers, Sealed Gas Detectors, Cosmic Rays
\end{keyword}

\end{frontmatter}

\section{Introduction}\label{sec:Introduction}
Resistive Plate Chambers (RPC), like most gaseous detectors, rely on the purity of the gas to keep their performance stable. For this reason, they operate with recirculation, purification and cleaning systems that attempt to keep the gas purity unchanged. In the first instance, gas degradation primarily occurs due to leaks/permeability in the system that allow atmospheric gases and/or humidity to enter, as well as the release of hydrofluorocarbons (HFCs), with C$_2$H$_2$F$_4$ being the main constituent. Additionally, molecular dissociation products created in gas amplification may also contribute to degradation over time. This results in the need to inject substantial amounts of fresh gas. The use of HFCs, particularly their release into the environment, has become a significant concern. This is due to the high Global Warming Potential (GWP) of HFCs, which contributes significantly to global warming. In fact, the European Union (EU) decreed the phase-out of HFCs in 2015. This poses serious problems for existing RPC systems, but especially for new systems, which will inevitably require alternative planning. In addition, the gas system associated with these detectors introduces considerable complexity and cost.

A possible solution to the problem, from an environmental perspective, although the complexity of the gas system would remain, is the replacement of these gases by others with a much lower GWP, the so-called green gases \cite{Man21}, with HFO-$1234$ze (C$_3$H$_2$F$_4$) as the most promising alternative \cite{GARILLOT2024169104}. 

Another possible solution would be to construct and operate RPCs without any gas supply, i.e. RPCs that contain gas but are hermetically sealed after construction, similar to the popular Geiger-M{\"u}ller tube. This concept has been baptized as sealed RPCs (sRPC) \cite{Lopes_2020, Blanco2023}. It would mitigate the problem of HFCs phase-out by drastically minimizing (by orders of magnitude) the amount of gas used today, thus minimizing its environmental impact to negligible levels. It would also eliminate any dependence on complex gas systems, allowing the expansion of this type of technology towards Cosmic Ray (CR) experiments through the construction of large, high-performance arrays at low cost, which would replace the simple Cherenkov water tanks in remote and difficult-to-access locations typical of CR experiments \cite{LOPES2023168446}.

In this work we present the characterization of a sealed RPC plane exposed to an intense radiation source for approximately one month, with the aim of verifying, in an accelerated manner, gas degradation effects induced by gas amplification phenomena that could compromise the operation of the detector. The results show no apparent degradation in any of the parameters studied suggesting the feasibility of applying this technology in Cosmic Ray experiments for extended periods of time.

\section{Experimental setup}\label{sec:setup}
\subsection{Sealed RPC}\label{sec:sealed}
\label{subsec:sealed}
\begin{figure}[h]%
	\centering
	\includegraphics[width=0.5\textwidth]{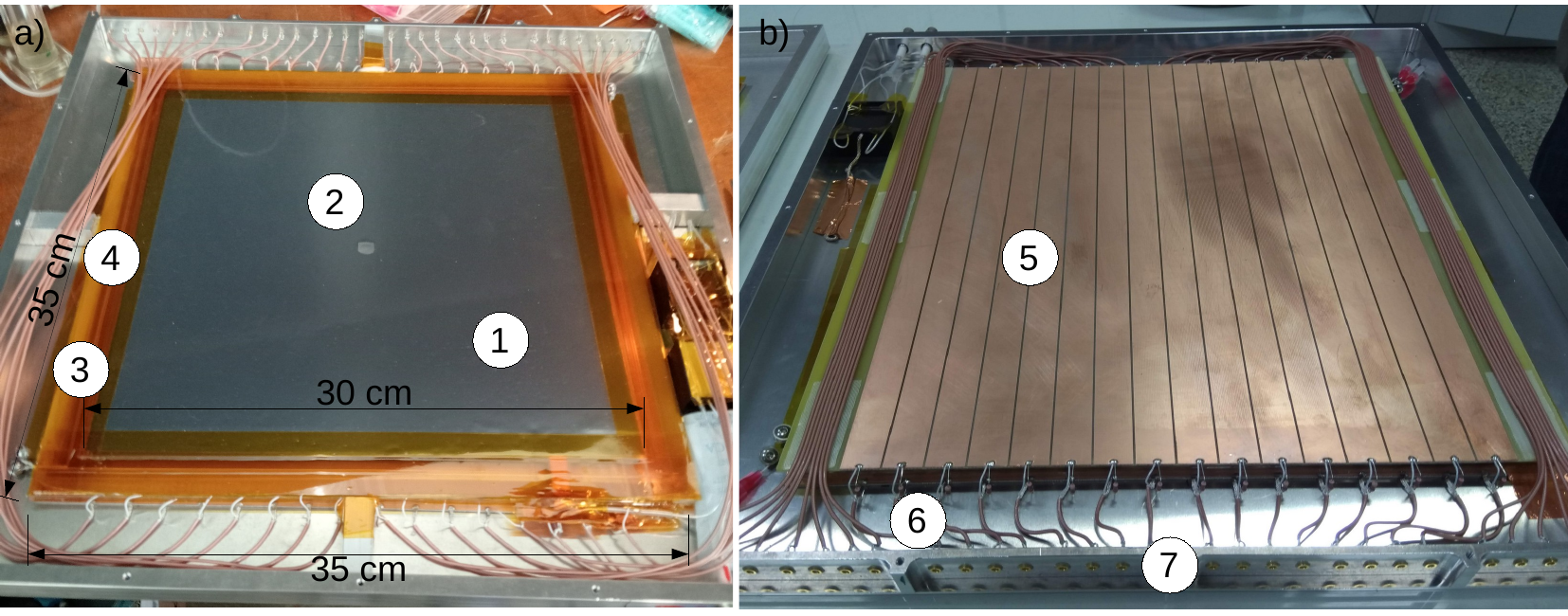}
	\caption{a) sRPC module showing: 1- HV layer, 2- Circular spacer in the center of the active area, 3- Strip spacer all around the periphery and 4- Mylar\textsuperscript{TM} and  Kapton\textsuperscript{TM} layers. b) sRPC plane showing: 5- Readout strip plane, 6- Coaxial cables and  7- MMCX RF feedthrough connectors.}\label{fig:setupSRPC}
\end{figure}

The sRPC module is a multi-gap structure \cite{CERRONZEBALLOS1996132} equipped with two $1$~mm gas gaps defined by three $2$~mm thick soda-lime glass electrodes \footnote{with bulk resistivity of $\approx 5x10^{12}$~$\Omega$cm at $25$~$^\circ$C} of about $350$~x~$350$~mm$^2$ separated by spacers. Each gap includes a circular spacer made of a soda lime glass disk ($10$~mm diameter) placed in the center of the active area and a strip ($25$~mm width), also made of soda lime glass, all around the periphery. For the assembly, the multi-gap structure is kept under controlled pressure and all peripheral surfaces are covered with an epoxy glue for sealing and mechanical strength. Gas inlets and outlets, one per gas gap, used for gas filling are made with standard plastic dispensing needles (from Loctite \textsuperscript{TM}). The High Voltage (HV) electrodes are made up of a semi-conductive\footnote{Based on an artistic acrylic paint with around $100~M\varOmega/\Box$.} layer airbrushed onto the outer surface of the outermost glasses of the multi-gap structure in an area of $300$~x~$300$~mm$^2$ and covered with a Mylar\textsuperscript{TM} and Kapton\textsuperscript{TM} layers for electrical insulation, see figure \ref{fig:setupSRPC}.a for reference. After assembly, the gaps are kept in gas flow with a mixture of $97.5$\%  C$_{2}$H$_{2}$F$_{4}$ and $2.5$\% SF$_{6}$ for a few days, after which the feedthroughs are closed.

The sRPC module is read out by a strip plane\footnote{Made of $1.6$~mm thick Flame Retardant 4 (FR4) Printed Circuit Board (PCB).} equipped, on one side, with sixteen $18$~mm wide, $19$~mm pitch and $400$~mm long copper strips, located on top of the module. A ground plane, placed on bottom completes the readout structure. The entire structure is enclosed in an aluminum box that provides the necessary electromagnetic insulation and mechanical rigidity. Every strip is connected at both ends to a coaxial cable inner conductor, while the outer conductor is connected to the ground plane. The coaxial cables are connected to MMCX RF feedthrough connectors, see figure \ref{fig:setupSRPC}.b for reference. 

\subsection{Plastic scintillator mini-telescope}\label{sec:telescope}
In order to trigger on atmospheric muons, a set of four $40$~x~$40$~x~$10$~mm$^3$ plastic scintillators read out on the smaller face by one Silicon Photomultiplier (SiPM), MICROFC-60035-SMT, are stacked vertically to form a muon mini-telescope. The scintillators are positioned such that the solid angle subtended by the outermost scintillators always remains within the active area of the sRPC.

\begin{figure}
\centering
\includegraphics[width=0.5\textwidth]{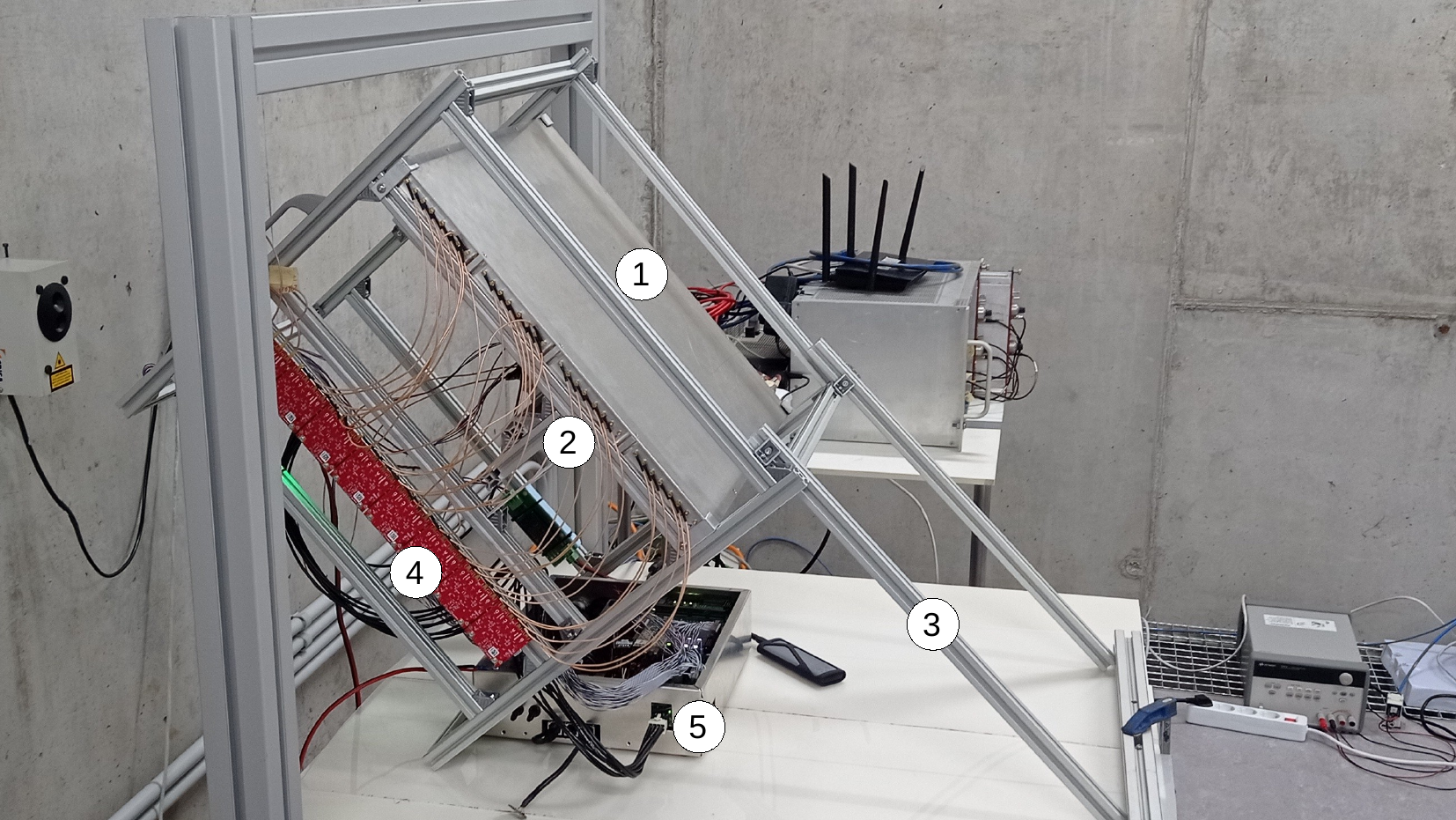}
\caption{Bunker arrangement: 1- sRPC, 2- location of the mini-telescope based on four plastic scintillators (not visible on the picture), 3- aluminum structure, 4- FEE and 5- DAQ box containing the TDC board and a single board computer together with the necessary ancillary elements.}
\label{fig:setup}       
\end{figure}

\subsection{Electronics}\label{sec:electronics}
Twenty-eight of the thirty-two signals from the sRPC (two strips are not measured to allow the reading of the SiPM signals), along with the four SiPM signals, are fed into fast Front-End Electronics (fFEE)\cite{HADES_FEE} borrowed from the HADES RPC-TOF detector \cite{HADES_RPC} capable of measuring time and the fast (electronic) component of the RPC signal charge in a single channel. The resulting digital signals are read out by the a TRB3sc board \cite{TRB3sc} equipped with $32$ multi-hit TDC (TDC-in-FPGA technology) channels with a time precision better than $20$~ps and a multipurpose logic unit. This logic unit allows the triggering either on muons passing through the four plastic scintillators, and thus study the response of the sRPC, or on the OR of all the signals coming from the sRPC, identified from now on as plastic scintillator and self-trigger respectively. The TDC board, along with a single-board computer and the necessary ancillary elements, such as, the low-voltage power supplies, temperature, pressure, and relative humidity sensors, are housed in a compact and portable box. A complete description of this DAQ box can be found in reference \cite{mingo}. 

With this arrangement we are able to provide information on:
\begin{itemize}
	\item Charge, $Q$, as the sum of the induced charge on the strips (or the mean charge, $<Q>$, as the mean of the $Q$ distribution), and streamer probability as the percentage of events with a charge higher than $1.2$ pC.
	\item Time, $T$, as the half-sum of the times at both ends for the strip with maximum charge.
	\item Longitudinal position to the strips, $Y$, as the difference of times at both ends for the strip with maximum charge multiplied by the propagation velocity on the strip, evaluated to $165$ mm/ns.
	\item Transversal position to the strips, $X$, as the position of the strip with maximum charge. Both, longitudinal and transversal positions are determined with a resolution better than $10$ mm sigma. 
    \item Charge and time for the four scintillators. 
\end{itemize}

For the irradiation, all of the above elements where installed in front of a cobalt source in a bunker. The arrangement can be seen in figure \ref{fig:setup}. The sRPC is directly exposed to the cobalt source, and the mini-telescope is located at the rear of it. The entire assembly was tilted at approximately $45^\circ$ to maximize the flux of atmospheric muons into the mini-telescope. In this way it is possible to calculate the efficiency of the sRPC by simply dividing the number of muons seen by the sRPC by the muons triggered by the mini-telescope.

\section{Results}\label{sec:results}
\begin{figure}[h!!!]
\centering
\includegraphics[width=0.5\textwidth]{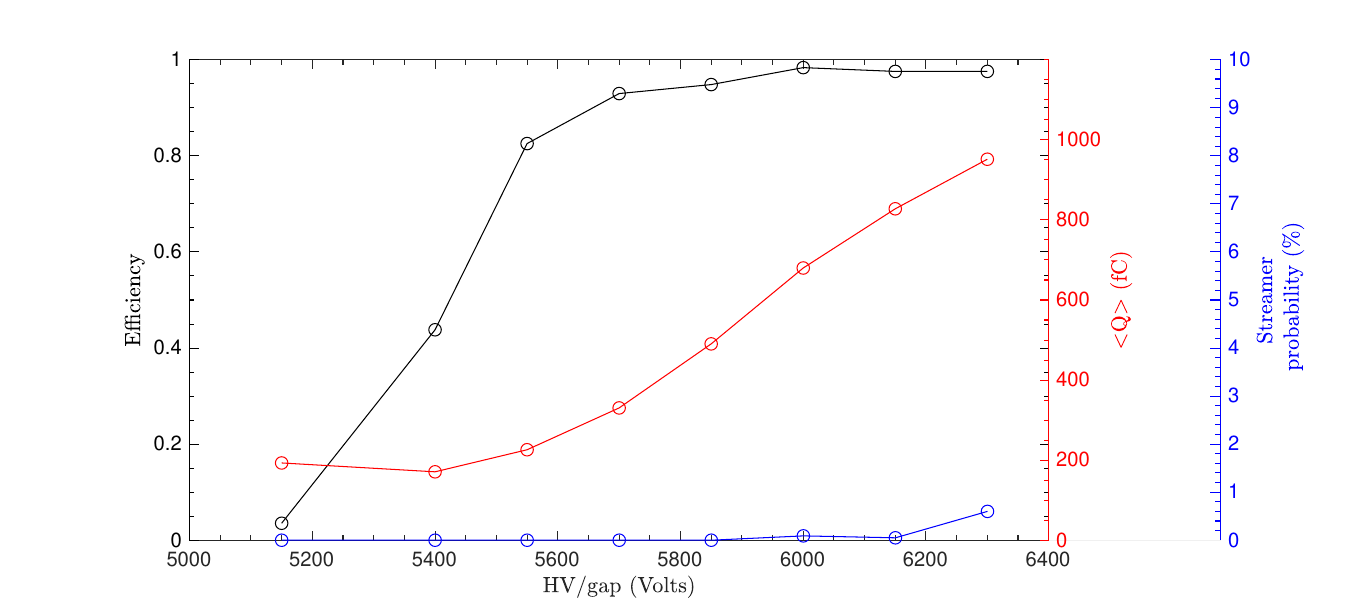}
\caption{Response of the sRPC to atmospheric muons before bunker irradiation as a function of the voltage applied in each gap showing an efficiency plateau of around $98\%$ above $6000$~V.}
\label{fig:HVScan}       
\end{figure}

Before irradiation in the bunker, the sRPC was characterized in the laboratory. Figure \ref{fig:HVScan}
shows the efficiency, average charge and streamer probability as a function of the applied voltage in each gap. The efficiency plateau appears above $6000$~V showing an average values of $97.9\%$ along with an average charge higher than $600$~fC and a streamer probability of less than 1\% at $6300$~V. Based on this plot the HV working point was set to $6000$~V.

\begin{figure}
\centering
\includegraphics[width=0.5\textwidth]{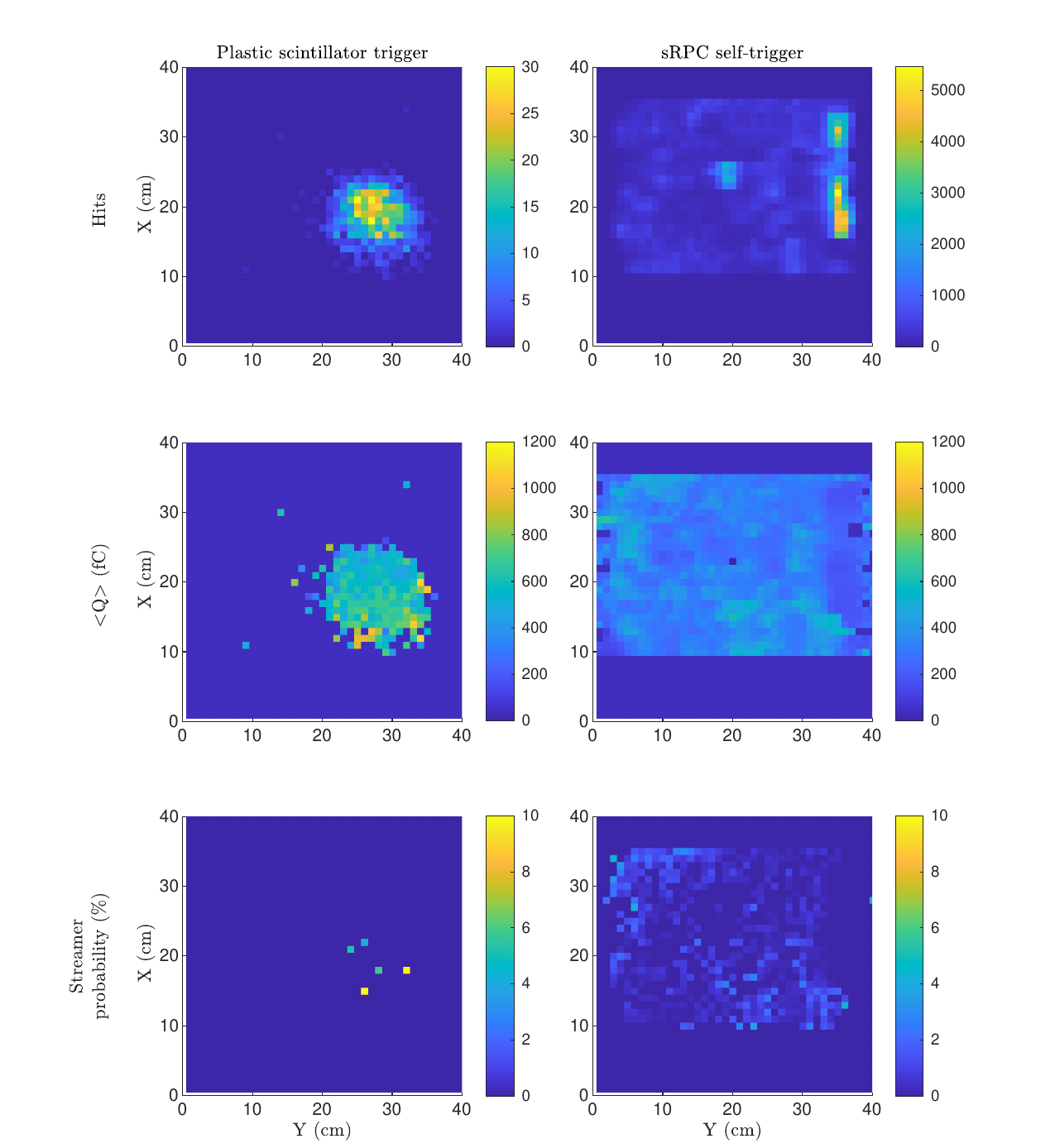}
\caption{Response of the sRPC before bunker irradiation as a function of the 2D position. a) b) and c) correspond to atmospheric muons and d), e) and f) to self-trigger.}
\label{fig:XYScan}       
\end{figure}

Figure \ref{fig:XYScan} shows the hits, average charge and streamer probability as a function of X and Y for a HV of $6000$~V. The left column, figures \ref{fig:XYScan}.a, b and c, correspond to the trigger generated by the plastic scintillators. It is possible to clearly identify the spot in the middle of the sRPC created by the atmospheric muons with a uniform average charge and a very low, almost non-existent, level of streamer probability. The right column, plots \ref{fig:XYScan}.c, d and e, correspond to self-trigger. Figure \ref{fig:XYScan}.d shows an accumulation of hits in the center and right side, probably associated with hot spots related to the spacers. Figure \ref{fig:XYScan}.e shows the average charge with no noticeable correlation with position and a modest streamers probability over the detector surface is shown in Figure \ref{fig:XYScan}.f. No modulation was observed depending on the position of the detector. 

\begin{figure}[h]
\centering
\includegraphics[width=0.5\textwidth]{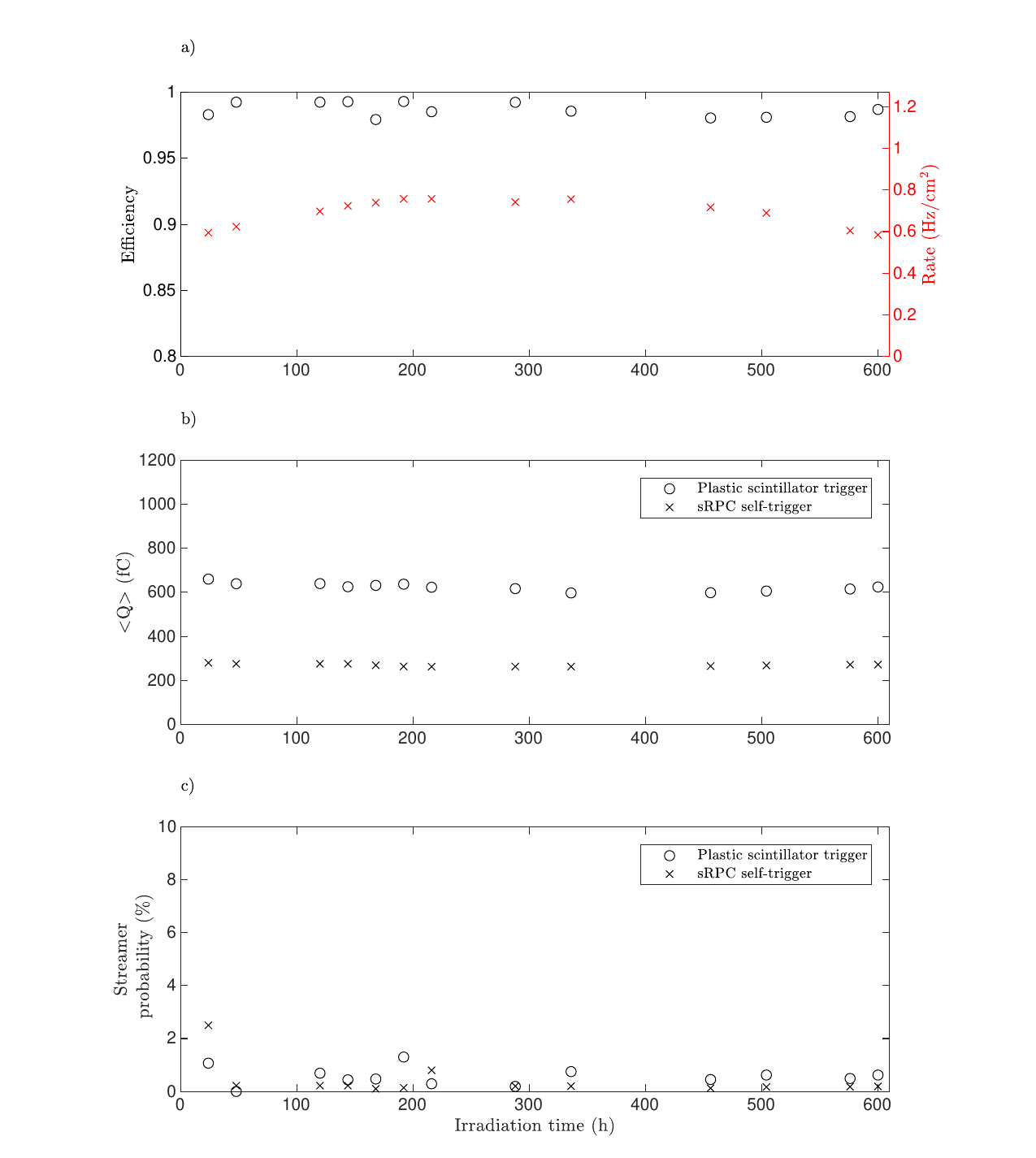}
\caption{Response of the sRPC to atmospheric muons during the irradiation period of approximately $600$ hours with the cobalt source. The analysis includes efficiency, background rate, average charge and streamer probability. Each point corresponds to approximately two hours of data collection without the cobalt source, after twenty two hours of exposure.}
\label{fig:Irradiation}       
\end{figure}

Originally the aim was to measure atmospheric muons in parallel with the photons emitted by the cobalt source in order to perform a continuous monitoring and characterization of the detector. In practice this was not possible due to the saturation of the sRPC. The cobalt source in its minimum irradiation configuration (including extra shielding) strongly saturates the sRPC with a visible rate of around $60$~Hz/cm$^2$ showing an average charge of around $180$~fC and consequently a low efficiency of around $16\%$. Finally, the sRPC was irradiated in periods of twenty two hours per day, stopping for two hours to measure the response to atmospheric muons. This is shown in figure \ref{fig:Irradiation}. Each point corresponds to approximately $2$ hours of data collection without cobalt source after $22$ hours of exposition to the cobalt source. The sRPC was irradiated under this configuration a total time of around $575$ hours ($24$ days).

Figure \ref{fig:Irradiation}.a shows the efficiency for cosmic muons and the rate (dark count rate) generated by the sRPC, figure \ref{fig:Irradiation}.b shows the average charge both for the plastic scintillator and self-trigger while figure \ref{fig:Irradiation}.c shows the streamer probability for the same triggers as a function of the irradiation time. No noticeable dependence on the irradiation time is observed. It can be concluded that the performance of the sRPC was not influenced by this irradiation time of $575$ hours. 

Considering the average charge and the observed rate in the sRPC, the integrated charge during irradiation ($0.018$~C) corresponds to approximately $60$ times the integrated charge without irradiation. Thus, the irradiation time in the bunker is equivalent to a 60-fold longer time without irradiation. For example, in a Cosmic Ray experiment, this would correspond to $1435$ days of operation. This result should be taken with some caution, as there is a significant difference between irradiation and non-irradiation periods. During irradiation periods, the sRPC is saturated and therefore operates at lower gain than nominal. This may cause some phenomena in the gas multiplication, which could affect the gas quality, to be absent or to contribute differently. However, from the point of view of the integrated charge, the calculation is correct.

\begin{figure}
\centering
\includegraphics[width=0.5\textwidth]{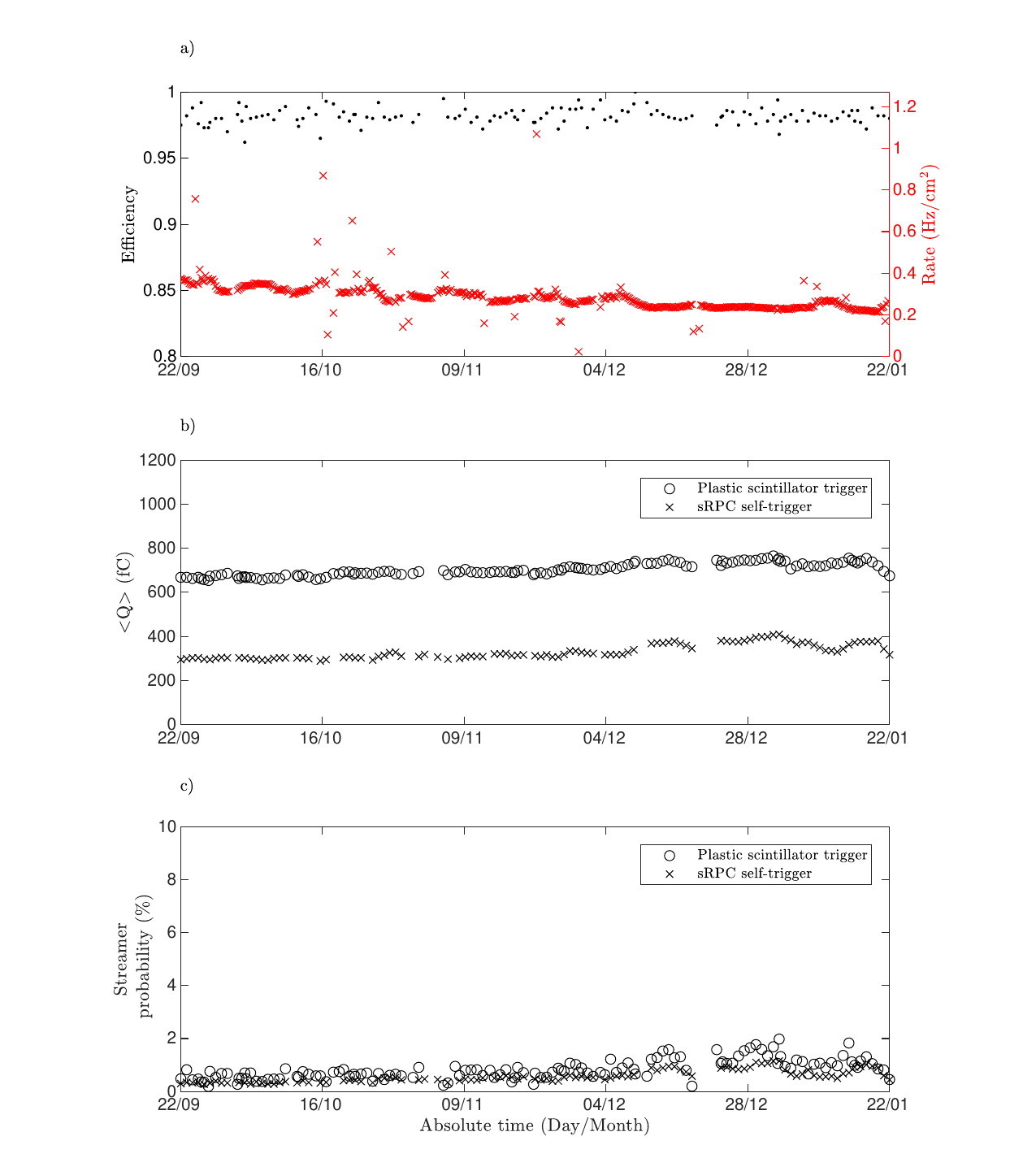}
\caption{Response of the sRPC to atmospheric muons during four months in the laboratory after bunker irradiation. The analysis includes efficiency, background rate, average load and streamer probability.}
\label{fig:longTerm}       
\end{figure}

In order to continue the evaluation of the longevity of the technology, the sRPC was reinstalled in the laboratory and operated for the last four months just after the irradiation in the bunker. Figure \ref{fig:longTerm} shows the same variables as in figure \ref{fig:Irradiation} for this 4-month period, where the sRPC has simply been exposed to natural Cosmic Ray radiation. Again, no relevant changes in any of the variables are observed with an average efficiency of $98.2\%$.

\section{Conclusions}\label{sec:conclusions}
A sRPC, a RPC requiring no continuous gas flow for its operation, with dimensions of $350$~x~$350$~mm$^2$, consisting of a multi-gap structure with two $1$~mm gaps was irradiated with an intense radioactive source for nearly one month with the aim of verifying, in an accelerated manner, gas degradation effects induced by gas amplification phenomena that could compromise the operation of the detector. Additionally, after irradiation, the sRPC was operated in the laboratory for four more months without being exposed to any radioactive source, only exposed to the natural muon flux and the background signals generated by the detector itself.

The results show no noticeable changes in the variables under study: efficiency, background rate, average charge and streamer probability after $24$ days of irradiation, equivalent to $1435$ days of operation without irradiation and even after $120$ days of operation in the lab following bunker irradiation. These results suggest the possibility of using sealed RPC technology in Cosmic Ray or other low counting rate experiments for long periods of time, without problems due to gas amplification phenomena affecting the gas quality.

\section{Acknowledgments}
This work was supported by Funda\c{c}\~ao para a Ci\^encia e Tecnologia, Portugal in the framework of the project CERN/FIS-INS/0006/2021 and 2024.00269.CERN. The authors would like to thank the hospitality and professionalism of the Radiophysics Laboratory at the University of Santiago de Compostela, where the irradiation was carried out, and especially to Dr. Diego M González-Castaño and his team.

\bibliographystyle{elsarticle-num-names}
\bibliography{paper-bibliography.bib}

\begin{thebibliography}{10}
\expandafter\ifx\csname url\endcsname\relax
  \def\url#1{\texttt{#1}}\fi
\expandafter\ifx\csname urlprefix\endcsname\relax\def\urlprefix{URL }\fi
\expandafter\ifx\csname href\endcsname\relax
  \def\href#1#2{#2} \def\path#1{#1}\fi

\bibitem{Man21}
B.~Mandelli,
  \href{https://indico.cern.ch/event/999799/contributions/4204191/attachments/2236047/3789965/BMandelli_ECFA.pdf}{Eco-gas
  mixtures and mitigation procedures for greenhouse gases (ghgs)} (2021).
\newline\urlprefix\url{https://indico.cern.ch/event/999799/contributions/4204191/attachments/2236047/3789965/BMandelli_ECFA.pdf}

\bibitem{GARILLOT2024169104}
G.~Garillot, Y.~Baek, D.~Hatzifotiadou, D.~Kim, J.~Kim, B.~Min, S.~Park,
  M.~Williams, R.~Zuyeuski,
  \href{https://www.sciencedirect.com/science/article/pii/S0168900224000305}{Operation
  of a low resistivity glass mrpc at high rate using ecological gas}, Nuclear
  Instruments and Methods in Physics Research Section A: Accelerators,
  Spectrometers, Detectors and Associated Equipment 1061 (2024) 169104.
\newblock \href {https://doi.org/https://doi.org/10.1016/j.nima.2024.169104}
  {\path{doi:https://doi.org/10.1016/j.nima.2024.169104}}.
\newline\urlprefix\url{https://www.sciencedirect.com/science/article/pii/S0168900224000305}

\bibitem{Lopes_2020}
L.~Lopes, P.~Assis, A.~Blanco, P.~Fonte, M.~Pimenta,
  \href{https://doi.org/10.1088/1748-0221/15/11/c11009}{Towards sealed
  resistive plate chambers}, Journal of Instrumentation 15~(11) (2020)
  C11009--C11009.
\newblock \href {https://doi.org/10.1088/1748-0221/15/11/c11009}
  {\path{doi:10.1088/1748-0221/15/11/c11009}}.
\newline\urlprefix\url{https://doi.org/10.1088/1748-0221/15/11/c11009}

\bibitem{Blanco2023}
A.~Blanco, P.~Fonte, L.~Lopes, M.~Pimenta,
  \href{https://doi.org/10.1140/epjp/s13360-023-04647-1}{Sealed (zero gas flow)
  resistive plate chambers}, The European Physical Journal Plus 138~(11) (2023)
  1021.
\newblock \href {https://doi.org/10.1140/epjp/s13360-023-04647-1}
  {\path{doi:10.1140/epjp/s13360-023-04647-1}}.
\newline\urlprefix\url{https://doi.org/10.1140/epjp/s13360-023-04647-1}

\bibitem{LOPES2023168446}
L.~Lopes, S.~Andringa, P.~Assis, A.~Blanco, N.~Carolino, M.~Cerda,
  F.~Clemêncio, R.~Conceição, O.~Cunha, C.~Dobrigkeit, M.~Ferreira,
  C.~Loureiro, L.~Mendes, J.~Nogueira, A.~Pereira, M.~Pimenta, J.~Saraiva,
  R.~Sarmento, P.~Teixeira, B.~Tomé,
  \href{https://www.sciencedirect.com/science/article/pii/S0168900223004369}{Outdoor
  systems performance and upgrade}, Nuclear Instruments and Methods in Physics
  Research Section A: Accelerators, Spectrometers, Detectors and Associated
  Equipment 1054 (2023) 168446.
\newblock \href {https://doi.org/https://doi.org/10.1016/j.nima.2023.168446}
  {\path{doi:https://doi.org/10.1016/j.nima.2023.168446}}.
\newline\urlprefix\url{https://www.sciencedirect.com/science/article/pii/S0168900223004369}

\bibitem{CERRONZEBALLOS1996132}
E.~C. Zeballos, I.~Crotty, D.~Hatzifotiadou, J.~L. Valverde], S.~Neupane,
  M.~Williams, A.~Zichichi,
  \href{http://www.sciencedirect.com/science/article/pii/0168900296001581}{A
  new type of resistive plate chamber: The multigap rpc}, Nuclear Instruments
  and Methods in Physics Research Section A: Accelerators, Spectrometers,
  Detectors and Associated Equipment 374~(1) (1996) 132 -- 135.
\newblock \href {https://doi.org/https://doi.org/10.1016/0168-9002(96)00158-1}
  {\path{doi:https://doi.org/10.1016/0168-9002(96)00158-1}}.
\newline\urlprefix\url{http://www.sciencedirect.com/science/article/pii/0168900296001581}

\bibitem{HADES_FEE}
D.~Belver, P.~Cabanelas, E.~Castro, J.~A. Garzón, A.~Gil, D.~Gonzalez-Diaz,
  W.~Koenig, M.~Traxler, Performance of the low-jitter high-gain/bandwidth
  front-end electronics of the hades trpc wall, IEEE Transactions on Nuclear
  Science 57~(5) (2010) 2848--2856.
\newblock \href {https://doi.org/10.1109/TNS.2010.2056928}
  {\path{doi:10.1109/TNS.2010.2056928}}.

\bibitem{HADES_RPC}
D.~Belver, A.~Blanco, P.~Cabanelas, N.~Carolino, E.~Castro, J.~Diaz, P.~Fonte,
  J.~Garzón, D.~Gonzalez-Diaz, A.~Gil, W.~Koenig, L.~Lopes, A.~Mangiarotti,
  O.~Oliveira, A.~Pereira, C.~Silva, C.~Sousa, M.~Zapata,
  \href{http://www.sciencedirect.com/science/article/pii/S0168900208019633}{The
  hades rpc inner tof wall}, Nuclear Instruments and Methods in Physics
  Research Section A: Accelerators, Spectrometers, Detectors and Associated
  Equipment 602~(3) (2009) 687 -- 690, proceedings of the 9th International
  Workshop on Resistive Plate Chambers and Related Detectors.
\newblock \href {https://doi.org/https://doi.org/10.1016/j.nima.2008.12.090}
  {\path{doi:https://doi.org/10.1016/j.nima.2008.12.090}}.
\newline\urlprefix\url{http://www.sciencedirect.com/science/article/pii/S0168900208019633}

\bibitem{TRB3sc}
T.~collaboration, \href{http://trb.gsi.de/}{Trb} (2025).
\newline\urlprefix\url{http://trb.gsi.de/}

\bibitem{mingo}
C.~Soneira-Landín,
  \href{https://indico.global/event/6191/contributions/51206/attachments/25745/44426/miniTRASGO%20-%20RPC2024.pdf}{The
  first minitrasgo cosmic ray telescope} (2024).
\newline\urlprefix\url{https://indico.global/event/6191/contributions/51206/attachments/25745/44426/miniTRASGO%20-%20RPC2024.pdf}

\end{thebibliography}
\end{document}